\documentclass[aps,prl,twocolumn,longbibliography,superscriptaddress]{revtex4-1}
\usepackage[latin1]{inputenc}
\usepackage[english]{babel}
\usepackage{bm,graphicx,graphics,amsmath,amssymb,epsfig,color}
\usepackage{dcolumn,lipsum}

\newcommand{\mH}{\mathcal{H}}

\newcommand{\mE}{\mathcal{E}}

\newcommand{\mP}{\textcolor{black}{P}}
\newcommand{\mQ}{\textcolor{black}{Q}}

\newcommand{\be}{\begin{equation}}
\newcommand{\ee}{\end{equation}}
\newcommand{\bea}{\begin{eqnarray}}
\newcommand{\eea}{\end{eqnarray}}
\newcommand{\bse}{\begin{subequations}}
\newcommand{\ese}{\end{subequations}}
\newcommand{\comment}[1]{}
\newcommand{\gcol}[1]{{\color{black} #1}}

\begin{document}

\title{Thermalization without chaos in harmonic systems}

\author{Niccol\'o Cocciaglia}
\affiliation{Dipartimento di Fisica, Universit\`a di Roma ``Sapienza'', Piazzale A. Moro 2, I-00185, Roma, Italy}

\author{Angelo Vulpiani}
\affiliation{Dipartimento di Fisica, Universit\`a di Roma ``Sapienza'', Piazzale A. Moro 2, I-00185, Roma, Italy}

\author{Giacomo Gradenigo}
\affiliation{Gran Sasso Science Institute, Viale F. Crispi 7, 67100 L'Aquila, Italy}
\affiliation{INFN-Laboratori Nazionali del Gran Sasso, Via G. Acitelli 22, 67100 Assergi (AQ), Italy}

\email{giacomo.gradenigo@gssi.it}


\begin{abstract}

  Recent numerical results showed that thermalization of Fourier modes
  is achieved in short time-scales in the Toda model, despite its
  integrability and the absence of chaos. Here we provide numerical
  evidence that the scenario according to which chaos is irrelevant
  for thermalization is realized even in the simplest of all classical
  integrable system: the harmonic chain. We study relaxation from an
  atypical condition given with respect to~\emph{random} modes,
  showing that a thermal state with equilibrium properties is attained
  in short times. Such a result is independent from the orthonormal
  base used to represent the chain state, provided it is random.
  
\end{abstract}

\maketitle

\section{Introduction}

Whether chaos is a necessary or just a sufficient condition in order
for many-body Hamiltonian systems to exhibit thermalization is an old
debate, to which a final assessment has not yet been
given~\cite{PG98,Z05,CFLV08}. A possibility, outlined with interesting
mathematical argument first by Khinchin in 1949~\cite{K49} and then by
Mazur and van der Linden~\cite{ML63}, is that basically the
applicability of the statistical mechanics description to a
macroscopic object relies on the number of its microscopic
constituents being very large, irrespective to details of their
dynamics. There are in fact examples showing that chaos is not
necessary to have good statistical features~\cite{MM60}, sometimes
being not even sufficient~\cite{LPRV87,FMV91}. It is in this spirit
that two of us recently investigated how an atypical initial condition
relaxes to equipartition in an integrable Hamiltonian system, the Toda
chain~\cite{BVG21}. It turned out that in order to see fast relaxation
to thermal equilibrium it is sufficient to consider appropriate
observables, for instance observables almost independent from the
conserved quantities of the integrable system~\cite{BVG21}. For what
concerns the Toda model at not too low energies this is for instance
the case of the Fourier modes harmonic energies, which thermalize
fast. On the other hand, one could say that thermalization of
Fourier modes in the Toda chain is specific to the choice of variables
and to its intricate relation with the Toda modes; in fact it is well
known that, if one considers the Toda modes, the thermodynamic of the
system is well described by the Generalized Gibbs
Ensemble~\cite{S20,VR16,dLM16,dVBdLM20}. The goal of the present work
is to give further and more stringent evidences in favour of the
scenario tested numerically in~\cite{BVG21} and at the basis of the
Khinchin approach~\cite{K49}: in the large-$N$ limit details of the
microscopic dynamics are not relevant for statistical mechanics.

In the present paper we discuss an alternative way to study the
behaviour of a system which is a textbook example: the harmonic
chain. Our goal is to show that the lack of thermalization is a
property \emph{specific} of the Fourier modes, in the sense that there
are infinitely many other choices of canonical variables to represent
the chain state and that using such variables one can see fast
relaxation to a thermal state. In order to complement the results
of~\cite{BVG21} we will present here the numerical evidence of two
important facts: 1) there are \emph{many} (ideally infinitely many)
independent sets of canonical variables with respect to which the
system shows thermalization and 2) the relaxation to a thermal state
is a large-$N$ effect. The fact that the validity of statistical
mechanics can be generally claimed for any system \emph{only} in the
large-$N$ limit is particularly relevant for integrable systems, for
which at small sizes the regular behaviour of the dynamics is
manifest. It is therefore definitely worth to recall the enunciate of
the Khinchin result, in order to grasp the relevance of the bound
imposed by the large-$N$ limit. Given that $f(q,p):\mathbb{R}^{2dN}
\rightarrow\mathbb{R}$ is an observable of our many-body classical
Hamiltonian system, with $\langle f \rangle$ and $\overline{f}$
respectively its statistical and time averages, if $f$ is a sum
function, namely something of the kind $f(q,p) = \sum_{i=1}^N
f_i(q_i,p_i)$ (in practice any sort of additive quantity which can be
measured as a function of small portions of the system), then under
quite general hypothesis~\cite{K49} one has that the probability to
have a difference between $\langle f \rangle$ and $\overline{f}$ small
in $N$ is itself small in $N$. By denoting as $\mathcal{P}$ this
probability, in formulas we have:
\begin{align}
\mathcal{P}\left( \bigg| \frac{\overline{f}-\langle f \rangle}{\langle
  f\rangle}\bigg| \geq \frac{C_1}{N^{1/4}} \right) ~<~
\frac{C_2}{N^{1/4}},
\end{align}
where $C_1$ and $C_2$ are $\mathcal{O}(1)$ constants which do not
depend on $N$. Without going in further details of the mathematical
results we can notice that its underlying message is in agreement with
our results: irrespective to the nature of the microscopic dynamics
ensemble and dynamical averages are for practical purposes equivalent
in the large-$N$ limit.


\section{Model}

Let us consider an harmonic chain with the following Hamiltonian
\begin{align}
  \mH(q,p) = \sum_{i=1}^N \frac{p_i^2}{2m} + \frac{k}{2} \sum_{i=0}^N
  (q_{i+1}-q_i)^2,
  \label{eq:HToda}
\end{align}
where we can set masses and the elastic coefficient of the springs to
$m=k=1$ and fixed boundary conditions $q_0=q_{N+1}=0$. The Hamiltonian
in Eq.~(\ref{eq:HToda}) can be easily put in a diagonal form with the
following change of variables
\begin{align}
  \mQ_k &= \sqrt{\frac{2}{N+1}}~\sum_{i=1}^N~q_i~\sin\left( \frac{\pi i k}{N+1}\right) \nonumber \\
  \mP_k &= \sqrt{\frac{2}{N+1}}~\sum_{i=1}^N~p_i~\sin\left( \frac{\pi i k}{N+1}\right).
\end{align}
By applying the above transformation, which is a canonical change of
coordinates, one gets:
\begin{align}
  \mH(\mQ,\mP) = \frac{1}{2} \sum_{k=1}^N (\mP^2_k + \omega_k^2 \mQ^2_k) 
  \label{eq:Fourier-Hamiltonian}
\end{align}
where
\begin{align}
 \omega_k=2\sin \left( \frac{\pi k}{2N+2} \right)
\end{align}
is the angular frequency of the $k$-th normal mode.  It is then
convenient for our purposes to introduce the semi-canonical complex
variables:
\begin{align}
  z_k &= \frac{\mP_k + i \omega_k \mQ_k}{\sqrt{2\omega_k}} \nonumber \\
  z^*_k &= \frac{\mP_k - i \omega_k \mQ_k}{\sqrt{2\omega_k}},
\end{align}
such that
\begin{align}
  \mathcal{H}(z,z^*) &=  \sum_{k=1}^N \omega_k~ |z_k| ^2 \nonumber \\
  \lbrace z^*_k, z_q \rbrace  &= i ~\delta_{k,q} 
\end{align}

\section{Random Modes Thermalization}

The considered system is such that if only one harmonic is excited at
the beginning, i.e., one sets as initial condition $|z_k|^2 = E $, for
a given $k$ and $|z_q|^2 = 0$ for all $q \neq k$, energy is never
shared among Fourier modes due to the lack of interaction. In other
words, Fourier modes are integrability Liouville-Arnol'd theorem
variables, so that no signature of thermalization can be found by
definition looking at such variables. But let us consider a different
point of view. The harmonic chain is a system characterized by a set
of $2N$ real coordinate-momenta variables, or by a set of $N$ complex
coordinates, which can be chosen at will. The complex coordinates
$z_k$ represent just one of the infinitely many choices available.
Any random rotation in the $N$-dimensional complex space leads to
another possible system of coordinates. Let us denote with the symbol
\begin{align}
\theta = \lbrace \theta_1,\ldots,\theta_N \rbrace,
\end{align}
the $N$ random angles which define a random rotation in $\mathbb{C}^N$
and with $M(\theta)$ the unitary matrix, $M(\theta)\in U(N)$, which
represent this rotation. In particular we have
\begin{align}
M^\dagger M ~=~ M M^\dagger ~=~ {\bf 1},
\end{align}
where $M^\dagger$ means transpose and complex conjugate. We thus have
infinitely many choices of \emph{``random modes''}, which define as
random rotations of the Fourier modes and indicate as:
\begin{align}
z_k(\theta) = \sum_{q=1}^N M_{kq}(\theta)~z_q.
\end{align}
\begin{widetext}
It is easy to check that the random rotation does not alter the
Poisson parentheses structure between the variables:
\begin{align}
  \left\lbrace z_p(\theta), z^*_q(\theta) \right\rbrace &=
  \sum_{ij} ~M_{pi}(\theta)~M^*_{qj}(\theta) ~\left\lbrace
  z_i, z^*_j \right\rbrace = ~i ~\sum_{i=1}^N
  ~M_{pi}(\theta)~M^\dagger_{iq}(\theta) = i~\delta_{pq}
\end{align}
\begin{figure}
  \includegraphics[width=0.45\columnwidth]{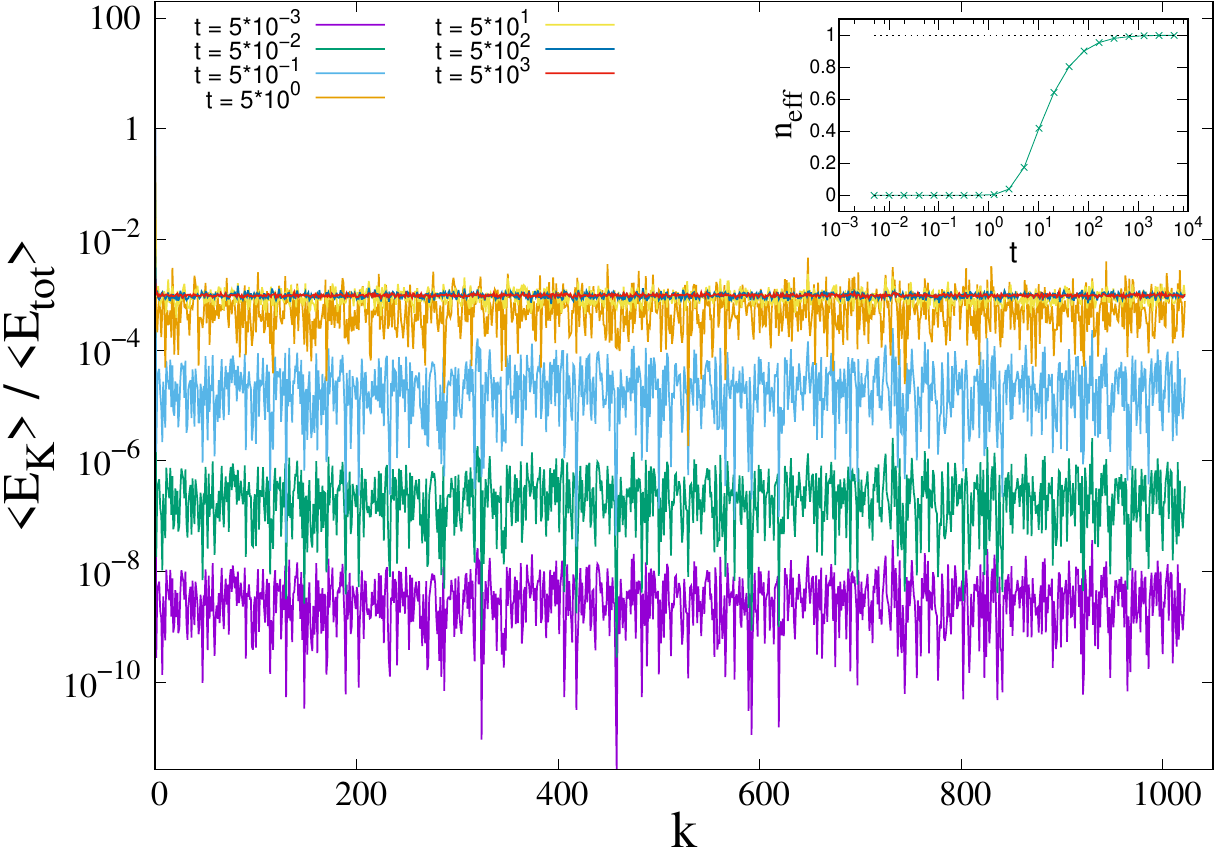}
  \includegraphics[width=0.45\columnwidth]{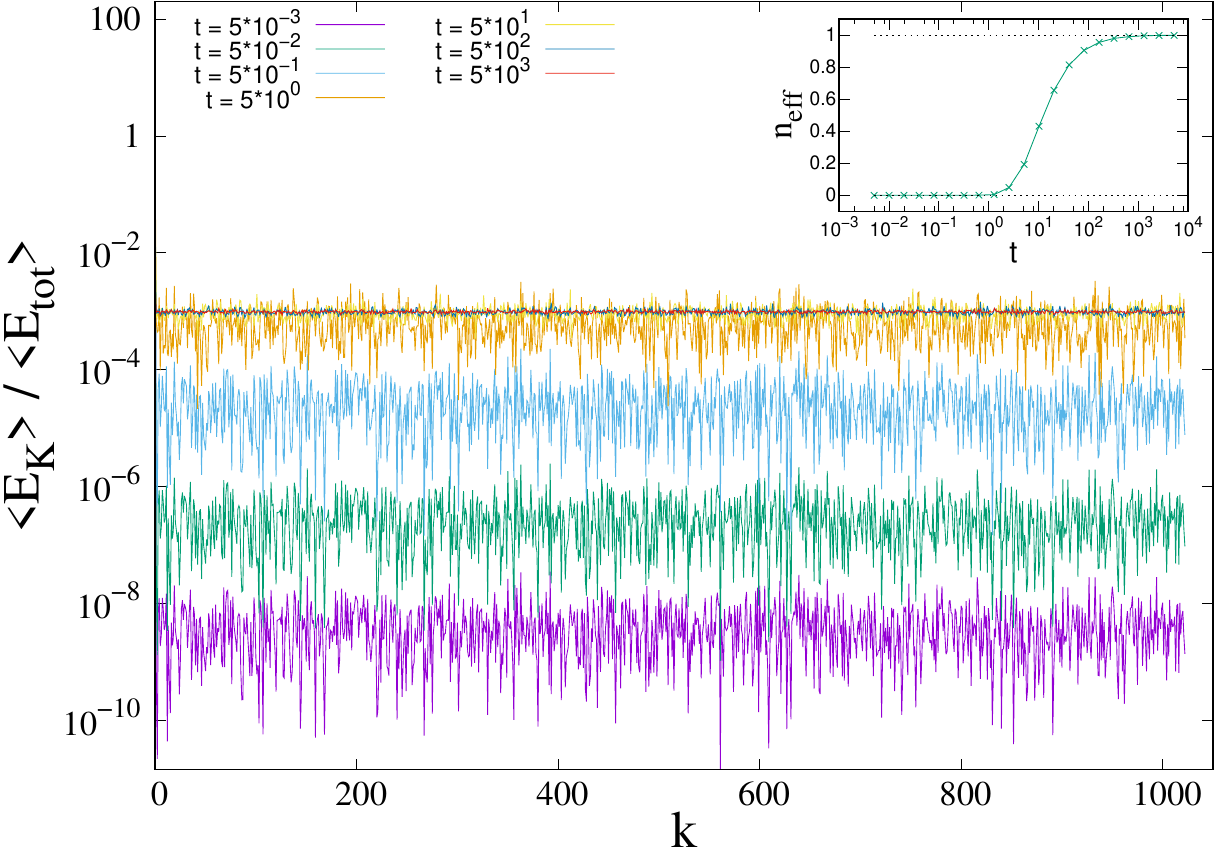}

  \caption{Panels $a)$ and $b)$ show two different choices of
    \emph{random modes}, i.e., two independent choices of the unitary
    random matrix $M(\theta)$. \emph{Main}: Normalized random modes
    energy spectrum at different times, $u_k(\theta,t)$ vs $k$ for
    increasing values of $t$; \emph{Inset}: effective number of
    degrees of freedom as a function of time, $n_{eff}(\theta,t)$ vs
    $t$.}
  
\label{fig1}
\end{figure}
\end{widetext}
Clearly, in terms of these random modes the Hamiltonian is not anymore
diagonal:
\begin{align}
\mathcal{H}(z(\theta),z^*(\theta)) =  \sum_{qp=1}^N c_{qp}(\theta)~z_q^*(\theta)~z_p(\theta), 
\end{align}
where
\begin{align}
c_{qp} = \sum_{k=1}^N
\omega_k~\mathcal{M}^*_{kq}(\theta)~\mathcal{M}_{kp}(\theta),
\label{eq:cqp-def}
\end{align}
with $\mathcal{M}(\theta) = [M^{-1}(\theta)]$, $\mathcal{M}(\theta)\in U(N)$.\\

We then define the energy of the random mode $k$ as
\begin{align}
\mathcal{E}_k(\theta) =  c_{kk}(\theta) |z_k(\theta)|^2,
\end{align}
since $c_{kk}(\theta)$ is real then also $\mathcal{E}_k(\theta)$ is
real. We have in fact by definition $c_{kk}(\theta) =
c_{kk}^*(\theta)$, see Eq.~(\ref{eq:cqp-def}).
We can thus write the Hamiltonian by separating the diagonal from the
non-diagonal contribution as
\begin{align}
  \mathcal{H}(z(\theta),z^*(\theta)) = \sum_{k=1}^N \mathcal{E}_k(\theta)  + \sum_{q\neq p}^N c_{qp}(\theta)~z_q^*(\theta)~z_p(\theta),
  \label{eq:Random-Fourier-Hamiltonian}
\end{align}
and then perform the standard FPUT numerical experiment, which amounts
to the numerical study of the relaxational dynamics when energy is
initially fed to a single random mode:
\begin{equation}
  \mathcal{E}_k =
  \begin{cases}
  E_0~~~~~~k=1\\   
  0~~~~~~~~k\neq 1\,.
  \end{cases}
  \label{eq:init-cond}
\end{equation}
Clearly the value of $E_0$ set in the initial condition does not
corresponds to the total energy of the system, which comprises the
non-diagonal terms of Eq.~(\ref{eq:Random-Fourier-Hamiltonian}). The
configuration of the system is then evolved according to the standard
Hamiltonian dynamics in the particles coordinate and momenta
representation.\\ In agreement with the traditional approach to the
Fermi-Pasta-Ulam-Tsingou problem~\cite{FPU55,LPRSV85,G05,FIK05,BCP13}
we have monitored along the dynamics the behaviour of the energy
spectrum and of the spectral entropy, which quantifies with a number
between zero and one the degree of energy equipartition between modes.
A way to quantify the degree of equipartition between modes is
obtained by monitoring the energy density
\begin{equation}
 u_k(\theta,t)=\frac{\langle \mathcal{E}_k(\theta) \rangle_t}{\langle \mathcal{E}_{tot}(\theta) \rangle_t}\,,
\end{equation}
where $\langle \mathcal{E}_{tot}(\theta) \rangle_t = \sum_{k=1}^N \langle
\mathcal{E}_k(\theta)\rangle_t$ and we consider cumulative time
averages of the kind
\begin{equation}
  \langle \mathcal{E}_k(\theta) \rangle_t  = \frac{1}{t} \int_0^t ds~\mathcal{E}_k(\theta,s) \,.
  \end{equation}
The same information can be represented by studying the behaviour of
the effective number of degrees of freedom $n_{\text{eff}}(\theta,t)$,
\begin{align}
  n_{\text{eff}}(\theta,t) = \frac{\exp\left(S_{\text{sp}}(\theta,t)\right)}{N}. 
\end{align}
where $S_{\text{sp}}(\theta,t)$ is the so-called \emph{spectral
  entropy}:
\begin{align}
  S_{\text{sp}}(\theta,t)  &= - \sum_{k=1}^N u_k(\theta,t)\log u_k(\theta,t) 
\end{align}
The parametric dependence of the spectrum $u_K(\theta,t)$ on the time
variable $t$ is illustrated in Fig.~(\ref{fig1}) for two different
choices of the random rotation $\theta$. Two remarkable observations
are in order: equipartition among random modes is reached quite fast
and this happens on the same time-scale for all choices of the random
rotation $\theta$. This means that, apart the Fourier basis, fast
relaxation to a thermal state is the \emph{typical} phenomenon for
random choice of the basis to represent the chain configuration. The
inset of each of the four panels of Fig.\ref{fig1} represents the
behaviour of the effective number of degrees of freedom
$n_{\text{eff}}(t)$: the typical sigmoidal shape with a fast
convergence to $n_{\text{eff}}(t)\approx 1$ signals the reaching of
equipartition. Despite integrability thermalization is observed even
in the harmonic chain, provided the \emph{``right variables''} are
considered. And, what is most remarkable, \emph{thermalization is
  typical}, while the lack of it is specific only to the
representation of the chain configuration in Fourier space. \gcol{The
  idea that in the large-$N$ limit the relevant thermodynamic
  properties of a system cannot be tight to a particular choice of
  coordinates was at the basis of an averaging strategy proposed
  already in~\cite{MPR94} to capture the essential thermodynamic
  features of the low temperature glassy phase of an optimization
  problem.}

\begin{figure}
  \includegraphics[width=0.95\columnwidth]{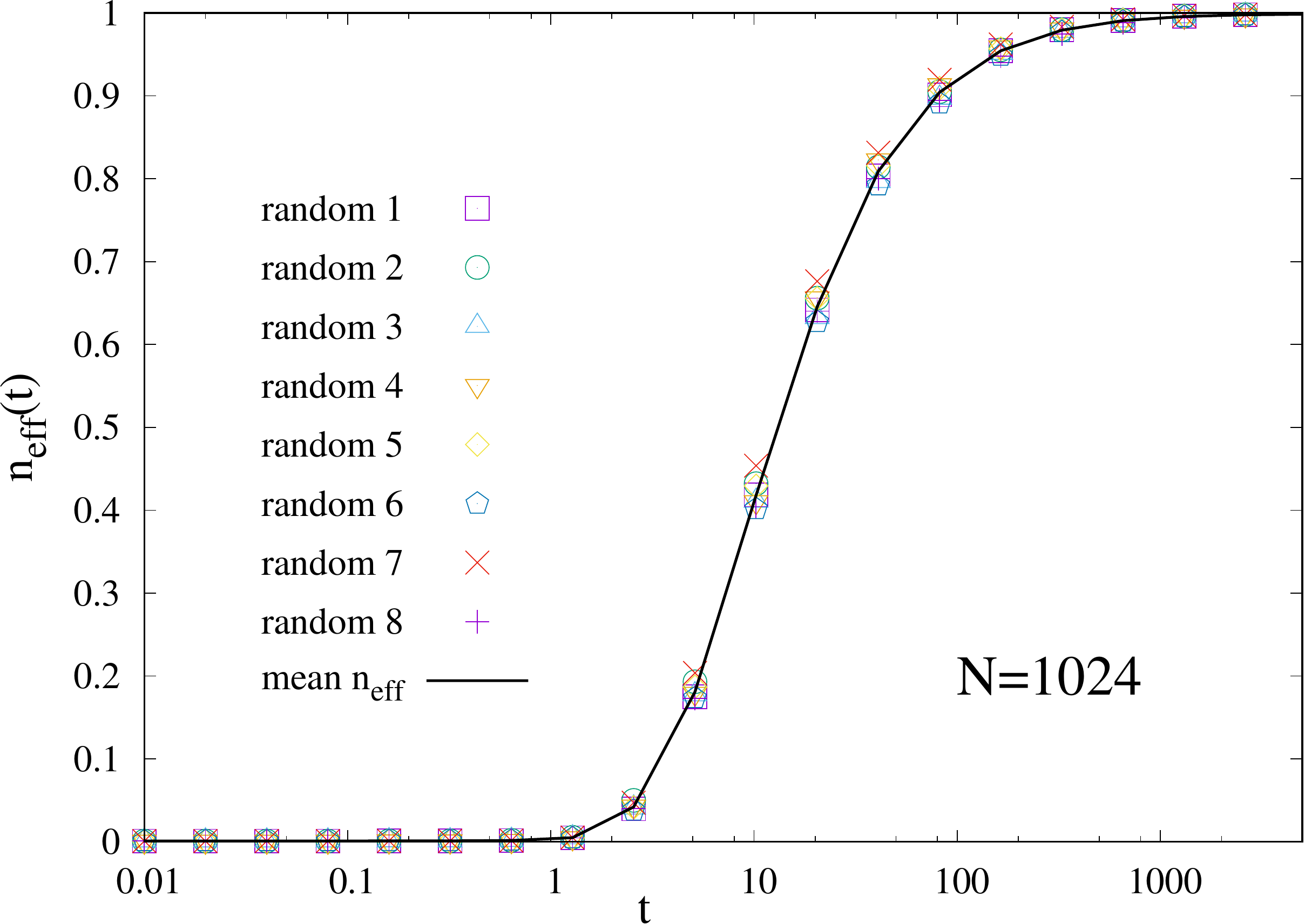}
  \caption{Effective number of degrees of freedom as a function of
    time, $n_{eff}(\theta,t)$ vs $t$: the continuous (black) line
    represents the average $\overline{n_{eff}}(t)$ over $30$ choices
    of the random modes base, points represent the individual
    behaviour of $8$ different instances.}
\label{fig2}
\end{figure}

In Fig.~\ref{fig2} the behaviour of the effective number of degrees of
freedom averaged over $M=30$ instances of the random modes,
$\overline{n_{\text{eff}}}(\theta,t)$ (continuous line), is compared
with the behaviour for single instances of the random modes,
$n_{\text{eff}}(\theta,t)$ (points): there is a clear evidence that
for a chain of $N=1024$ the behaviour of $n_{\text{eff}}(\theta,t)$
for each single instance is typical.

\begin{align}
  \overline{n_{\text{eff}}}(t) &= \frac{1}{M} \sum_{i=1}^M n_{\text{eff}}(\theta_i,t) \nonumber \\
\end{align}

\section{Equilibrium dynamics: importance of the large-$N$ limit.}

\begin{figure}
  \includegraphics[width=0.95\columnwidth]{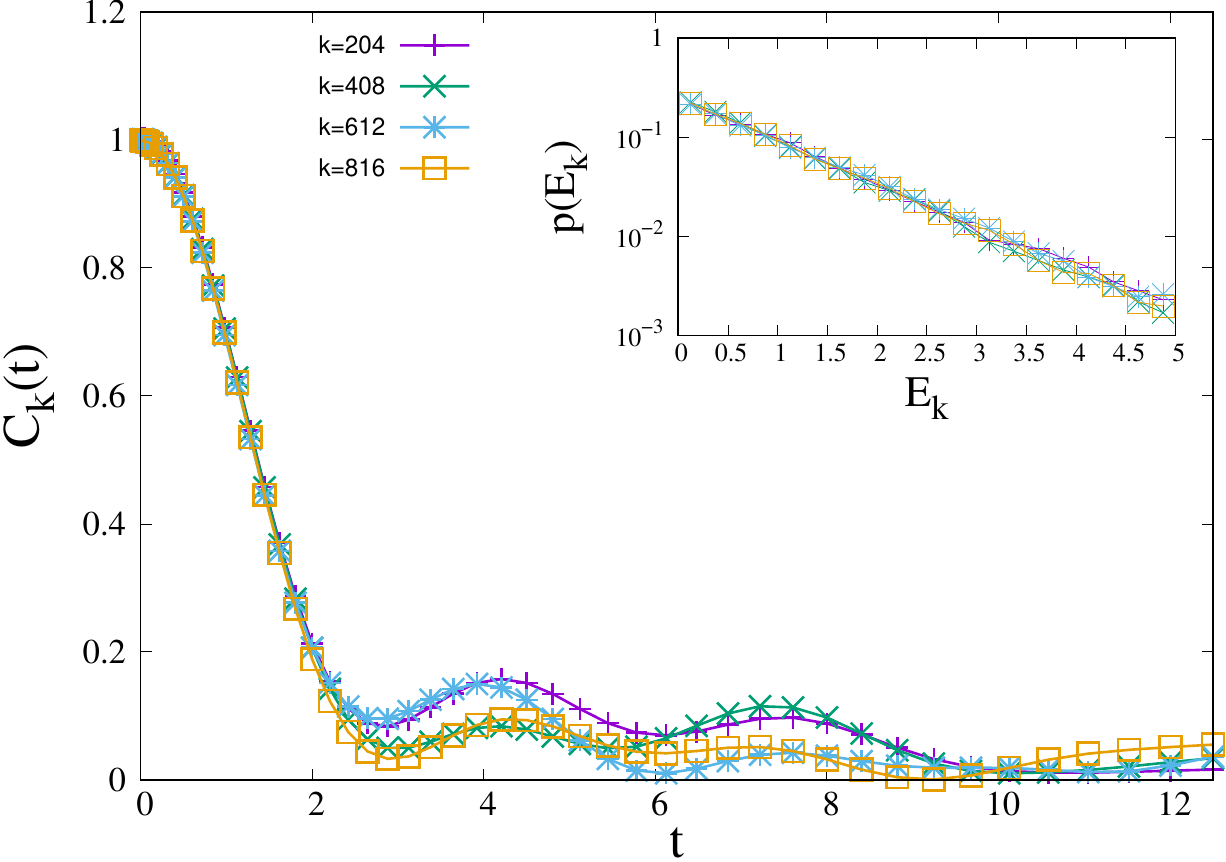}
    \caption{\emph{Main}: time auto-correlation function $C_k(t)$ as a
    function of time for different choices of the random modes index
    $k$, system size$N=1024$; \emph{Inset}: probability distribution
    of random-modes energy $\mathcal{E}_k$ for some values of $k$,
    system size $N=1024$.}
\label{fig3}
\end{figure}

Beside the ``standard'' FPUT-like numerical experiment discussed in
the previous section, probing the relaxation from an atypical initial
condition to a thermal state, it is important to investigate the
equilibrium dynamics of the system. The first test of thermalization
has been the study of the single-mode energy probability
distributions, simply defined as the histogram of values taken by
$\mathcal{E}_k$. In the inset of Fig.~\ref{fig3} it is shown that
$p(\mathcal{E}_k)$, plotted for different values of $k$, has in all
cases a nice exponential behaviour, as expected from a thermal
ensemble:
\begin{align}
  p(\mathcal{E}_k) \sim \exp\left( -b \mathcal{E}_k\right),
  \label{eq:pek}
\end{align}

where the value of $b$, proportional to the inverse of energy per
degree of freedom, is identical for all random modes $k$. The second
important test of thermalization is to check that the system reached a
stationary state where, \emph{``to all practical purposes''}, no
memory of a generic initial condition is left.

Clearly, this corresponds to a very pragmatic point of view: we look
at a generic random modes basis and simply ignore that the system is
integrable. How does look like an time auto-correlation function $C_k(t)$ of the
energy on the random mode $k$? Let us define $C_k(t)$ as
\begin{align}
 C_k(t) = \frac{\langle \mathcal{E}_k(t) \mathcal{E}_k(0)\rangle -
   \langle \mathcal{E}_k\rangle^2}{\langle
   \mathcal{E}_k^2\rangle-\langle \mathcal{E}_k\rangle^2}\,.
 \label{eq:ckt}
\end{align}
In the main panel of Fig.~\ref{fig3} it is shown the behaviour of
$C_k(t)$ for different $k$: the function rapidly decays to values
close to zero, around which it then oscillates. How it is possible
that an integrable system, moreover characterized by simple linear
interactions, presents such good signatures of an equilibrium
behaviour? Isn't it inconsistent with integrability? The way out to
this apparent contradiction comes by taking into account the same two
points which are the cornerstone of the Khinchin theorem: to consider
the appropriate observables and to consider the large-$N$ limit.
The dynamics of Fourier modes
amplitudes in the harmonic chain is clearly analytically predictable,
as done in a very insightful paper of Mazur and
Montroll~\cite{MM60}. 

Already in~\cite{MM60} in fact it was shown that, even in an
integrable system like an harmonic chain, if one takes into account a
``generic'' observable like the energy of a single particle of the
chain, the time-autocorrelation function decays fast to a zero. In
that case it was then shown that the imprint of the integrable nature
of the system was brought by a sort of quasi-periodic oscillations
around zero of amplitude amplitude $1/N^{1/2}$~\cite{MM60}. The
picture emerging already from~\cite{MM60} and corroborated by the
present results on random modes is therefore clear: if one considers
as observables functions which are a (random) combination of
\emph{all} the coordinates diagonalizing the Hamiltonian, the
``integrability effects'' are of order $\mathcal{O}(N^{-1/2})$. From
this point of view we can already find in the work of Mazur and
Montroll~\cite{MM60} the idea that thermalization comes about simply
with the thermodynamic limit and does not depend on the details of the
microscopic dynamics. For what concerns the present work we have shown
that the autocorrelation function of a random mode energy is for all
practical purposes equivalent to that of single particle in the
harmonic chain, since they both result from the combination of all
Fourier modes. In agreement with the results of~\cite{MM60}, it is
worth showing that even in the present case the fast decay of $\langle
\mE_k(t) \mE_k(0) \rangle$ is a large-$N$ effect. Without going
through a too much detailed finite-size analysis, we show here in
Fig.~\ref{fig4} that for small system sizes (e.g., $N=7$ particles)
the recurrent nature of the integrable dynamics clearly emerges.

\begin{widetext}

  In Fig.~\ref{fig4} is shown the time dependence of the
  autocorrelation function defined in Eq.~(\ref{eq:ckt}) for an
  harmonic chain of $N=7$ particles and for some values of $k$: it can
  be clearly seen that at a characteristic time $\tau\approx4520$ the
  system is back to its initial condition.

\begin{figure}
  \includegraphics[width=0.45\columnwidth]{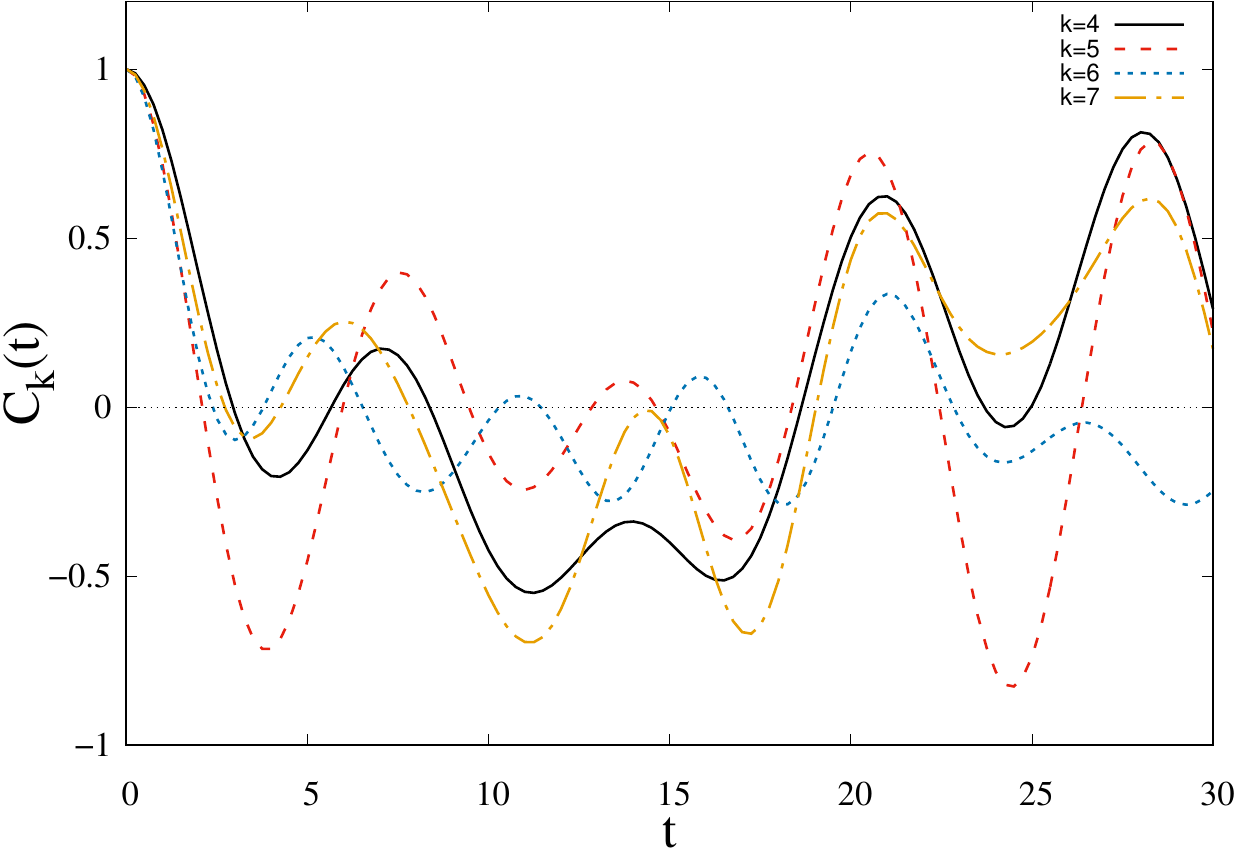}
  \includegraphics[width=0.45\columnwidth]{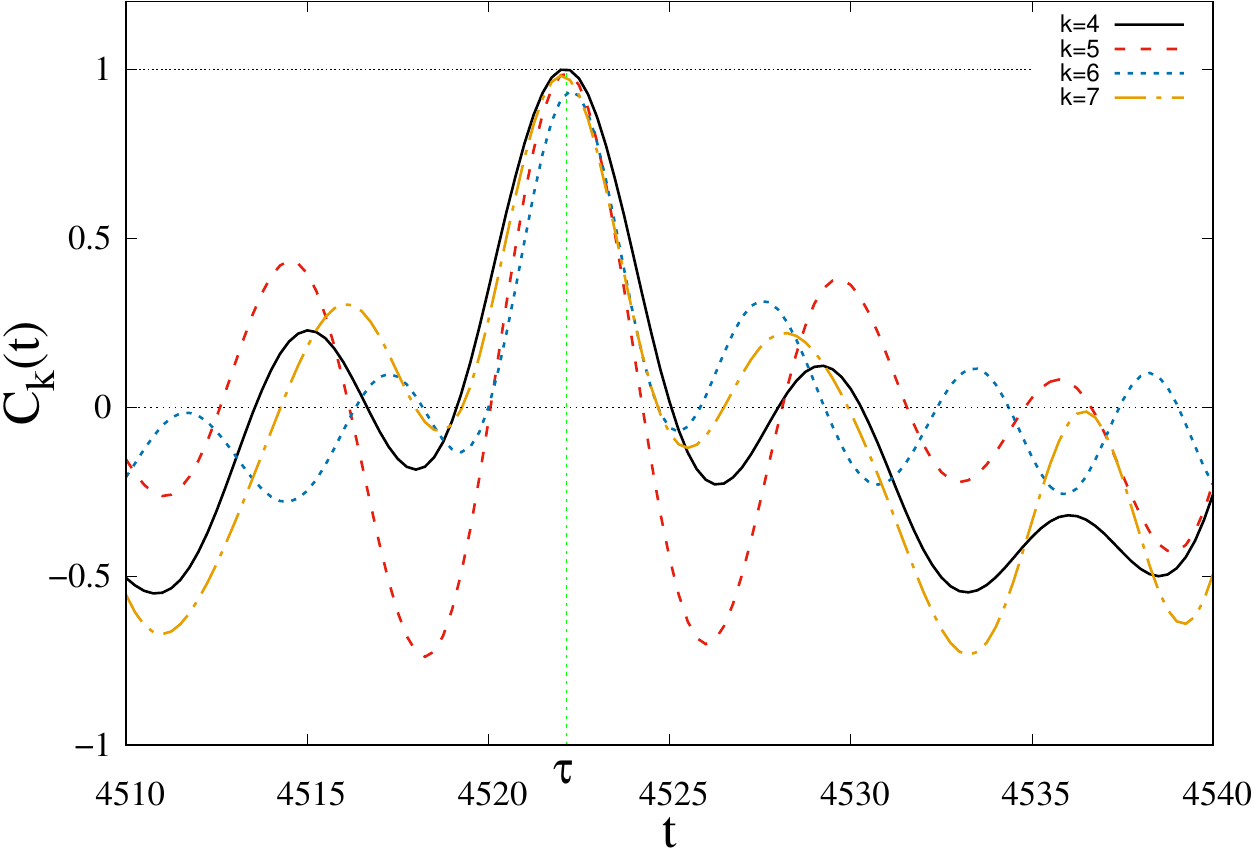}
  \caption{Time autocorrelation function $C_k(t)$ as a function of
      time for different choices of the random-modes index $k$, system
      size $N=7$.}
\label{fig4}
\end{figure}
\end{widetext}

\section{Overlaps distribution}
\label{overlaps}

Let us now represent in a quantitatively precise manner the statement,
already stressed by some of us in~\cite{BVG21}, that the variables
allowing to see thermalization in an integrable system must be
\emph{almost uncorrelated} to those diagonalizing the Hamiltonian. To
this aim we introduce the \emph{overlap} between Fourier modes and
random modes,
\begin{align}
|q_{kp}| = \bigg|\frac{\langle z_k(\theta) | z_q \rangle}{\langle z_q | z_q \rangle}\bigg| = |M_{kq}|,
\end{align}
which turns out to be nothing but the the matrix element $M_{kq}$ of
the random unitary matrix connecting Fourier modes and random modes
(we have to consider its modulus because it is a complex
variable). Since $|q_{kp}|=|M_{kq}|$, from the properties of random
unitary matrices the probability distribution $\rho(|q|)$ can be computed
exactly. For the ease of the reader we report this elementary large
deviation estimate in the Appendix, quoting here only the final
result:
\begin{align}
  \rho(|q|) = 2N |q| \exp\left( -N |q|^2\right)
  \label{eq:probq-mod}
\end{align}
By recalling that we are dealing with a complex variable,
$q\in\mathbb{C}$, the linear dependence on $|q|$ in front $\rho(|q|)$
has to be understood simply as the Jacobian element for the
representation of the complex variable in polar coordinates, so that
the probability distribution of the overlap,
$\mathcal{P}(q):\mathbb{C}\rightarrow\mathbb{R}$ is simply Gaussian:
\begin{align}
\mathcal{P}(q) \propto \exp\left( -N q^2\right),
\end{align}
\begin{figure}
  \includegraphics[width=0.95\columnwidth]{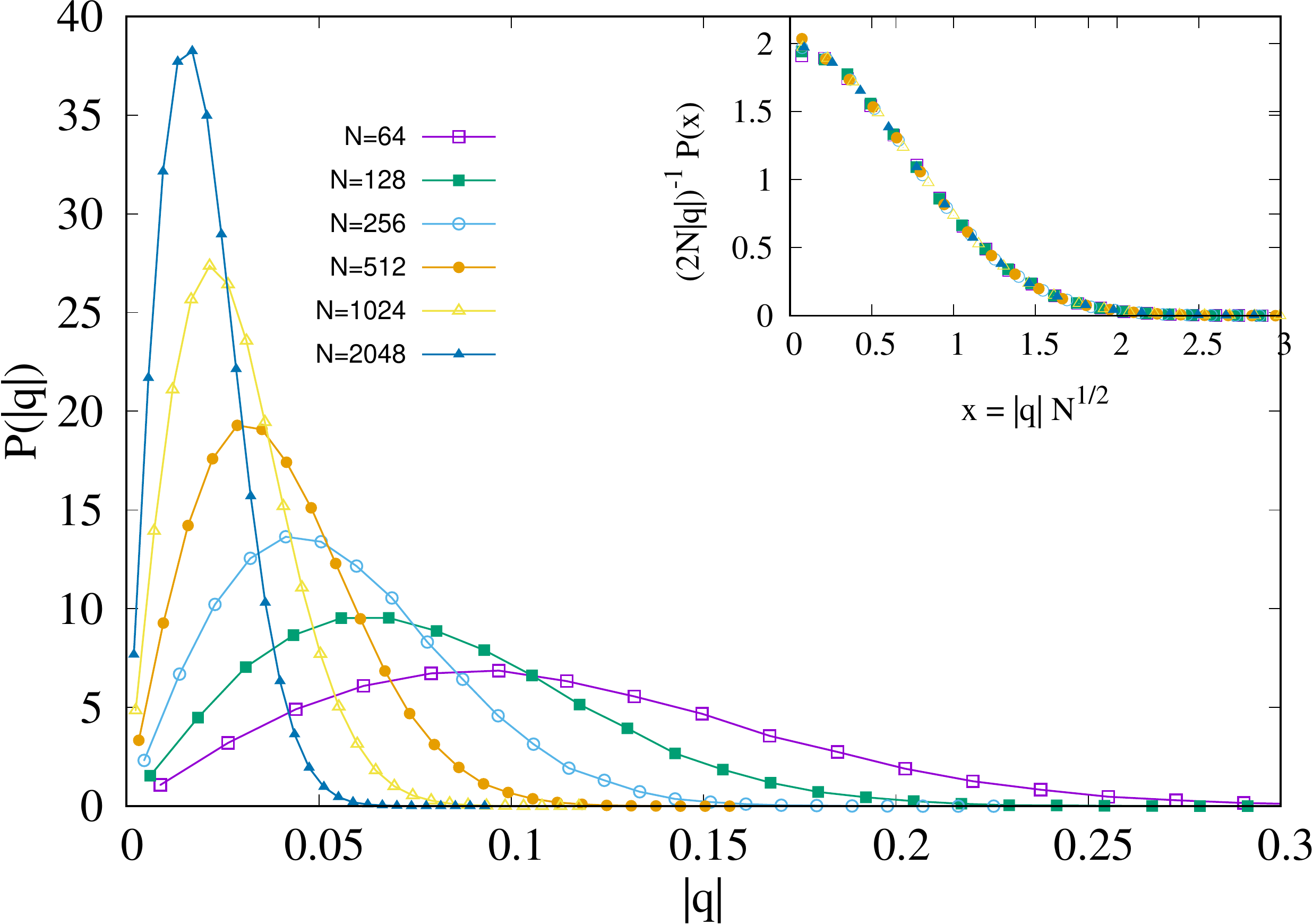}
  \caption{\emph{Main}: probability distribution $P(|q|)$ of the
    absolute value of the overlap between a Fourier mode and a random
    mode, for different sizes $N$ of the system; \emph{Inset}:
    collapsed data for different $N$.}
\label{fig5}
\end{figure}
which completes our discussion about choosing a set of canonical
coordinates which are \emph{random combination} of the coordinates
which make the integrable nature of the system manifest. In addition
to what already observed in~\cite{BVG21}, in the present work provided
sufficient evidence from data in favour of the hypothesis that the set
of (semi) canonical coordinates which behaves ``thermally'' are
infinitely many, actually all the random unitary transformations in
$M(\theta):\mathbb{C}^N\rightarrow\mathbb{C}^N$. It is remarkable how
the large-$N$ prediction of Eq.~(\ref{eq:probq-mod}) is in good
agreement with the numerical estimates of the probability distribution
of overlaps at finite $N$, which is shown in Fig.~\ref{fig5}. The
existence of an infinite number of basis of orthonormal functions to
represent the chain configuration, each base looking equally random to
any other, and the relevance of this property for the possibility to
detect equilibrium with respect to each base closely resembles a
similar property studied for the choice of eigenfunction basis in
quantum systems~\cite{GLTZ17}.

\section{Conclusions}
\label{conclusions}

The results presented in this work aims at clarifying a foundational
problem in statistical mechanics: is dynamical chaos really a
necessary condition to guarantee the thermalization of systems with
many degrees of freedom? Our results generalized and strengthned the
conclusions drawn for the Toda model in~\cite{BVG21}: thermalization
is achieved even in an integrable system, provided that an appropriate
choice of the observables is made. The study of the harmonic chain
added an insight: we have shown that there are many (ideally
infinitely many) different choices of canonical coordinates which
allow to detect thermalization in the harmonic chain. This evidence
tells us that the lack of equilibrium is really a specific property of
the coordinates which diagonalizes the Hamiltonian. \gcol{Robust
  thermodynamic properties must be independent from the choice of the
  coordinates used to write the partition sum, an idea already
  exploited with success in replica calculations for ordered systems
  with a glassy phase~\cite{MPR94}. This idea, which is substantiated
  mathematically by the Khinchin approach and of which we presented
  numerical evidence in this paper, tells us that even for integrable
  systems there are no a-priori reasons to not expect the validity of
  equilibrium ensembles. This of course is not in contradiction with
  well established frameworks for integrable systems such as the
  Generalized Gibbs
  Ensemble~\cite{S20,VR16,dLM16,dVBdLM20,RSMS09,CLNPT18,BCLNPT19,BCLN20}, the
  latter being specific of particular choice of the canonical
  coordinates. At the same time it is clear that the possibility of
  unveil specific properties of the dynamics crucially depends on the
  choice of variables, in particular for system where ergodicity is
  spontaneously broken as spin glasses~\cite{MPV87,CKP97,CKM19,K21} or
  systems highly structured and intrinsically characterized by
  multiple length/time scales like macro-molecules~\cite{PDC09}.}\\

\gcol{As a final remark, let us notice that a generalization of the
  present results to quantum mechanics may provide further insights on
  the thermalization mechanism in quantum isolated systems. The point
  made by our findings on the \emph{irrelevance of chaos} for the
  thermalization of harmonic systems are in fact quite similar to the
  Von Neumann quantum ergodic theorem approach~\cite{JVN29}. According
  to the latter thermalization is in fact related to an appropriate
  choice of observables, rather than to the spectral properties of the
  Hamiltonian. The same scenario, where the choice of observables
  plays a central role, is the one underlying the Eigenstate
  Thermalization Hypothesis~\cite{Sr99,DKPR16}. Numerical experiments
  to better explore the analogies between the foundation of
  statistical mechanics in classical and quantum systems, for instance
  investigating the scope of thermal ensembles applicability in the
  presence of an external driving~\cite{PFK21},
  i.e. fluctuation-dissipation relations, are therefore in order. \\}


{\bf Acknowledgments}~We thank J. Kurchan, R. Livi, S. Pappalardi and
A. Ponno for useful discussions. A.V. acknowledge partial financial
support of project MIUR-PRIN2017 \textit{Coarse-grained description
  for non-equilibrium systems and transport phenomena (CO-NEST)}.

\newpage

\setcounter{section}{0}

\section{Appendix A}
\label{app-A}

In this Appendix we show how to compute the probability distribution
of the generic matrix element $M_{ki}$ (for simplicity of notation,
let us omit the angle $\boldsymbol{\theta}$). To this aim it is
possible to exploit the fact that each row/column of the matrix
represents one (normalized) element of an eigenvector basis of
$\mathbb{C}^N$, so that the following condition holds:

\begin{equation}
    \sum_{i=1}^N \lvert M_{ki} \lvert^2\ = A \ \ \ \ \ \forall k\ ,
    \label{normaliz}
\end{equation}

where $A$ is the normalization that we will fix to 1 at the end of the
calculation. Changing the notation so that: $\zeta_i = M_{ki}$, we
have that the volume spanned by the elements of a row in $M$ reads:

\begin{equation}
    \Omega_N(A) = \int \prod_{i=1}^N d\operatorname{Re}(\zeta_i) d\operatorname{Im}(\zeta_i)\ \delta \left( A - \sum_{i=1}^N \lvert \zeta_i \lvert^2 \right)\ ,
\end{equation}

so that the joint probability density is:

\begin{equation}
    \rho (\zeta_1, \dots, \zeta_N \lvert A) =
    \frac{1}{\Omega_N(A)}\ \delta \left( A - \sum_{i=1}^N \lvert
    \zeta_i \lvert^2 \right)\ .
\end{equation}

Since we are looking for the distribution of a single matrix element,
we are interested in the marginal:

\begin{equation}
    \rho (\zeta_1 \lvert A) = \int \prod_{i=2}^N
    d\operatorname{Re}(\zeta_i) d\operatorname{Im}(\zeta_i)\ \rho
    (\zeta_1, \dots, \zeta_N \lvert A)\ .
\end{equation}

In order to compute $\Omega_N(A)$, let us switch to polar coordinates
$\displaystyle \begin{cases} \operatorname{Re}(\zeta_i) = r_i
  \cos(\phi_i) \\ \operatorname{Im}(\zeta_i) = r_i \sin(\phi_i)
\end{cases}:$

\begin{equation}
    \Omega_N(A) = (2\pi)^N \int_0^{\infty} \left[ \prod_{i=1}^N dr_i\ r_i \right]\ \delta \left( A - \sum_{i=1}^N r_i^2 \right)\ .
    \label{onlydr}
\end{equation}

In order to get to the result, we compute its Laplace transform:

\begin{align}
  & \Omega_N(\mu) = \int_0^{\infty} dA\ e^{-\mu A}\ \Omega_N(A) \nonumber \\
  &= (2\pi)^N \int_0^{\infty} \left[ \prod_{i=1}^N dr_i\ r_i \right]\ \exp{[-\mu \sum\nolimits_i r_i^2]} = \nonumber \\
  &  = \left[ 2\pi \int_0^{\infty} dr\ r\ e^{-\mu r^2} \right]^N = \left( \frac{\pi}{\mu} \right)^N \nonumber \\
  & = \pi^N\ \exp{[-N\log\mu]}
\end{align}

and then its inverse transform by applying the saddle point
approximation, in the large-$N$ limit:

\begin{align}
  \Omega_N(A) & = \frac{1}{2\pi i} \int_{\mu_0 - i\infty}^{\mu_0 + i\infty} d\mu\ e^{\mu A}\ \Omega_N(\mu) \nonumber \\
  & = \frac{\pi^N}{2\pi i} \int_{\mu_0 - i\infty}^{\mu_0 + i\infty} d\mu\ e^{N[\mu a - \log \mu]} \nonumber \\
  & = \frac{\pi^N}{2\pi i}\ \exp{\left\{ N [\mu^*(a)a - \log\mu^*(a)] \right\}}\ ,
\end{align}

having put $a = A/N$ and defined $\mu^*(a)$ as the saddle point value,
which can be explicitly computed using the saddle point equation:

\begin{equation}
    \frac{\partial}{\partial \mu} \left[ \mu a - \log(\mu) \right] = 0 \ \ \ \Longrightarrow \ \ \ \mu^* = \frac{1}{a} = \frac{N}{A} = N\ ,
\end{equation}

having finally imposed the normalization condition: $A=1
\ \longrightarrow \ a = 1/N$. Since we selected the value of $\mu$
which corresponds to the correct normalization of eigenvectors, we can
simply compute the Laplace transform of the unnormalized marginal
$\rho (\zeta \lvert A)$ and plug in the correct value of $\mu^*$ in
order to estimate its leading behaviour. Namely, we first consider:

\begin{equation}
    \rho (r \lvert A) \propto \int_0^{\infty} \left[ \prod_{i=2}^N dr_i\ r_i \right]\ r\ \delta\left( A - r^2 - \sum_{i=2}^N r_i^2 \right)
\end{equation}

and then its Laplace transform, with the same procedure we already
employed:

\begin{equation}
    \rho (r \lvert \mu) \propto \theta(r)\ r e^{-\mu r^2} \ \ \ \ \ \stackrel{\mu^* = N}{\longrightarrow} \ \ \ \ \  \rho (r \lvert N) = \mathcal{N}~\theta(r)\ r e^{-N r^2}\ ,
\end{equation}

where $\theta(\cdot)$ is the Heaviside function and $\mathcal{N}$ the
normalization factor. Therefore, considering that $r_i = \lvert
\zeta_i \lvert = \lvert M_{ki} \lvert$, it's been shown that the
probability distribution of the overlap between one of the normal
modes and one of the random ones, in the large-$N$ limit, is the
product of a linear function and a Gaussian centered in $r=0$ with
standard deviation: $\displaystyle \sigma_g \sim 1 / \sqrt{N}$. We can
easily compute the normalization factor of the distribution, as well
as the mean value and the variance, by solving Gaussian integrals; the
results are:

\begin{align}
  & A = 2N \nonumber \\
  & \langle r \rangle =  \sqrt{\frac{\pi}{4N}} \nonumber \\
  & \sigma^2 = \langle r^2 \rangle - \langle r \rangle^2 = \left( 1 - \frac{\pi}{4} \right) \frac{1}{N} 
    \label{moments}
\end{align}

In conclusion, the probability distribution for the overlaps $\lvert q
\lvert$ (dropping the indices) reads:

\begin{equation}
    \rho (\lvert q \rvert) = 2 N \lvert q \rvert e^{-N \lvert q \rvert^2}
    \label{overlapdistr}
\end{equation}

\end{document}